\documentclass[conference]{IEEEtran}
\IEEEoverridecommandlockouts
\usepackage{multirow}
\usepackage{cite}
\usepackage{amsmath,amssymb,amsfonts}
\usepackage{algorithmic}
\usepackage{graphicx}
\usepackage{textcomp}
\usepackage{xcolor}
\usepackage{placeins}
\def\BibTeX{{\rm B\kern-.05em{\sc i\kern-.025em b}\kern-.08em
    T\kern-.1667em\lower.7ex\hbox{E}\kern-.125emX}}
\begin{document}

% \usepackage{balance}

% ============================
% TITLE
% ============================
\title{Self-Supervised Audio Representation Learning for Pediatric Asthma Detection in Emergency Care Using Digital Stethoscope Recordings}

% ============================
% AUTHORS
% ============================
\author{
\IEEEauthorblockN{
Fatemeh Bagheri\IEEEauthorrefmark{1}\IEEEauthorrefmark{2},
Thalia Pandolfi\IEEEauthorrefmark{3},
Ervin Sejdić\IEEEauthorrefmark{1}\IEEEauthorrefmark{2},
Rohit Mohindra\IEEEauthorrefmark{2}\IEEEauthorrefmark{4}\IEEEauthorrefmark{5}
}
\IEEEauthorblockA{
\IEEEauthorrefmark{1}Department of Electrical and Computer Engineering, University of Toronto, Toronto, Canada\\
\IEEEauthorrefmark{2}North York General Hospital, Toronto, Canada\\
\IEEEauthorrefmark{3}Temerty Faculty of Medicine, University of Toronto, Toronto, Canada
\\
\IEEEauthorrefmark{4}Division of Emergency Medicine and Institute of Medical Sciences, University of Toronto, Toronto, Canada\\
\IEEEauthorrefmark{5} Schwarz Reisman Emergency Medicine Institute, Toronto, Canada\\
Email: fatemeh.bagheri@mail.utoronto.ca, thalia.pandolfi@mail.utoronto.ca,  ervin.sejdic@utoronto.ca, rohit.mohindra@nygh.on.ca
}
\thanks{This study was supported by North York General Foundation - Innovation Fund.}
}

% \title{Self-Supervised Audio Representation Learning for Pediatric Asthma Detection in Emergency Care Using Digital Stethoscope Recordings\\

% \thanks{Identify applicable funding agency here. If none, delete this.}
% }

% \author{\IEEEauthorblockN{1\textsuperscript{st} Fatemeh Bagheri}
% \IEEEauthorblockA{\textit{dept. name of organization (of Aff.)} \\
% \textit{name of organization (of Aff.)}\\
% City, Country \\
% email address or ORCID}
% \and
% \IEEEauthorblockN{2\textsuperscript{nd} Thalia Pandolfi}
% \IEEEauthorblockA{\textit{dept. name of organization (of Aff.)} \\
% \textit{name of organization (of Aff.)}\\
% City, Country \\
% email address or ORCID}
% \and
% \IEEEauthorblockN{3\textsuperscript{rd} Ervin Sejdic}
% \IEEEauthorblockA{\textit{dept. name of organization (of Aff.)} \\
% \textit{name of organization (of Aff.)}\\
% City, Country \\
% email address or ORCID}
% \and
% \IEEEauthorblockN{4\textsuperscript{th} Rohit Mohindra}
% \IEEEauthorblockA{\textit{dept. name of organization (of Aff.)} \\
% \textit{name of organization (of Aff.)}\\
% City, Country \\
% email address or ORCID}

% }

\maketitle
\begin{center}
\footnotesize
Accepted for publication at the 2026 48th Annual International
Conference of the IEEE Engineering in Medicine and Biology Society
(EMBC 2026). © 2026 IEEE. Personal use of this material is permitted.
Permission from IEEE must be obtained for all other uses, in any
current or future media, including reprinting/republishing this
material for advertising or promotional purposes, creating new
collective works, for resale or redistribution to servers or lists,
or reuse of any copyrighted component of this work in other works.
\end{center}
\begin{abstract}
Accurate diagnosis of pediatric asthma in emergency departments remains challenging due to overlapping respiratory symptoms, time constraints, and the limited feasibility of pulmonary function testing in young children. This study investigates the feasibility of pediatric asthma detection in the emergency department using breath sound recordings and machine learning. Thirty-second breath sounds were collected from six chest locations in 31 pediatric patients (10 asthmatic,
21 non-asthmatic) and analyzed using pretrained self-supervised speech representation models (HuBERT, WavLM, and Wav2Vec 2.0) for feature extraction, with patient age and sex incorporated into the feature representations. Conventional machine learning classifiers were trained and evaluated using patient-level stratified group 5-fold cross-validation and leave-one-patient-out validation to ensure the generalizability of the findings. Among the evaluated approaches, Wav2Vec 2.0 combined with histogram-based gradient boosting achieved the strongest and most consistent performance, yielding an accuracy of 0.84, sensitivity of 0.80, specificity of 0.86, and F1-score of 0.76 under both evaluation protocols. The consistency of performance across validation strategies suggests promising generalization to unseen patients. These findings suggest that pretrained self-supervised audio representations offer a promising, non-invasive approach for pediatric asthma detection in real-world emergency department settings, where objective respiratory assessment is often limited.
\end{abstract}

\begin{IEEEkeywords}
Asthma detection, self-supervised learning, machine learning
\end{IEEEkeywords}
 % \vspace{0.5cm }
\section{Introduction}
Asthma is one of the most common chronic respiratory conditions in children \cite{21}. It is a frequent cause of pediatric emergency department (ED) visits for acute respiratory complaints \cite{1}. However, accurate diagnosis in the ED remains challenging, as the diagnostic gold standard (i.e., pulmonary function testing) is often impractical in emergency settings and difficult to perform in young children \cite{4}. As a result, clinicians frequently rely on subjective clinical judgment in environments where multiple pediatric respiratory conditions present with overlapping symptoms, which increases the risk of diagnostic error \cite{Yang2017Misdiagnosis,Aaron2018UnderOverDiagnosis}. These challenges are exacerbated in emergency departments by high patient volume, environmental noise, and time constraints \cite{Zorc2003MissedAsthma}. Under such conditions, studies have shown that the identification of pediatric asthma in the ED is limited, with reported sensitivities on the order of approximately 40-45\% when compared against reference standards or comprehensive clinical evaluation \cite{Sanders2007EDAsthma}.

%, all of which can compromise the reliability of auscultation-based assessment 
% Failure to recognize asthma during emergency visits has been associated with increased disease burden and a higher likelihood of subsequent ED utilization \cite{10,Zorc2003MissedAsthma}.

Previous studies have explored the use of machine learning (ML) methods to support asthma-related clinical decision-making, including diagnosis, risk stratification, and prediction of exacerbations \cite{11, 13, 14}. While these approaches demonstrate the potential of ML in asthma care, most of them rely on pulmonary function measurements, and existing audio-based studies have largely been conducted in controlled or non-acute settings, limiting their applicability to the dynamic, noisy, and time-constrained environment of emergency care \cite{12}.
%most of these approaches depend on pulmonary function measurements, and no study has evaluated respiratory sound–based models in the ED. Existing

Recent advances in self-supervised learning (SSL) have addressed several of these limitations by enabling the learning of high-level acoustic representations directly from raw audio without requiring large labeled datasets. SSL-based pretrained speech models have shown strong performance across a range of non-clinical tasks, such as emotion recognition \cite{17}, as well as clinical applications including respiratory disease screening (e.g., COVID-19 detection from cough and breath sounds) \cite{15,16}. However, these models have not been evaluated for pediatric asthma detection in emergency department settings, where recordings are acquired under noisy, time-constrained conditions and prior studies have largely relied on audio collected in controlled environments \cite{m11}. By leveraging large-scale unlabeled audio corpora during pretraining, SSL models can capture robust and noise-tolerant representations that are well suited for data-limited and acoustically challenging environments. Models such as HuBERT \cite{19}, WavLM \cite{18}, and Wav2Vec 2.0 \cite{20} exemplify this paradigm and offer a promising foundation for breath sound analysis in the ED.

In this work, we investigate the feasibility of pediatric asthma detection in the ED using breath sound recordings and ML-based analysis. Specifically, we examine whether pretrained SSL speech models can be leveraged for feature extraction from real-world ED recordings and whether these representations can support patient-level asthma classification using conventional ML classifiers, thereby addressing key limitations of prior studies that relied on controlled recording conditions and lacked evaluation in the ED environment.

 \vspace{0.5cm }
\section{Methods}
\begin{figure*}[htbp]
\centering
\includegraphics[width=2\columnwidth]{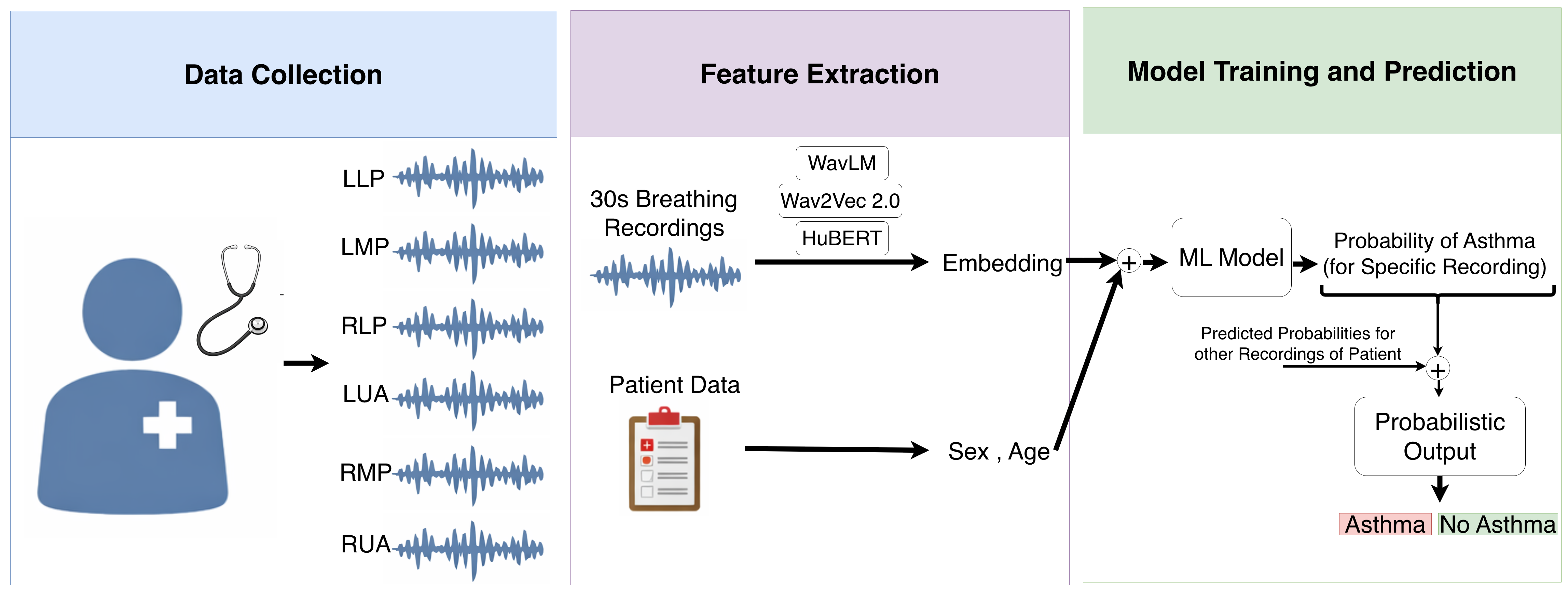}
\caption{Overview of study pipeline.}
\label{fig}
\end{figure*}

Figure \ref{fig} provides an overview of the study pipeline.

\subsection{Data collection}
Thirty-second breathing sound recordings were collected from six chest locations (i.e., left lower posterior (LLP), left mid posterior (LMP), right lower posterior (RLP), left upper anterior (LUA), right mid posterior (RMP), and right upper anterior (RUA)) using an electronic stethoscope in the ED at North York General Hospital, Toronto, Canada. Recordings were obtained from pediatric patients aged 1–18 years presenting with suspected asthma (n=31; 10 asthmatic, 21 non-asthmatic; 13 female; mean age 6.4 years). The final clinical diagnosis was independently adjudicated by two board-certified pulmonologists. The experimental procedures involving human subjects described in this paper were approved by the Institutional Review Board of University of Toronto and North York General Hospital.

\subsection{Feature extraction}
All breath sound recordings were resampled to a target sampling rate of 16 kHz to ensure compatibility with the pretrained models. For each 30-second recording, audio feature extraction was performed using pretrained self-supervised speech representation models, including HuBERT, WavLM, and Wav2Vec 2.0. Then, we applied temporal mean pooling over the last hidden layer to obtain one fixed-length embedding per recording. Next, patient age and sex were concatenated with the audio embeddings prior to classification. The resulting feature vectors were used as inputs to the ML models.

\subsection{Models}
Several ML models were evaluated for pediatric asthma detection using the extracted audio-based feature representations. The evaluated classifiers included logistic regression (LogReg), histogram-based gradient boosting (HistGB), random forest, and support vector machines. Results are reported for the top two performing models.
All models were trained using standardized feature inputs to ensure comparability across classifiers. To mitigate the effects of class imbalance, balanced class weights were applied where supported. Model hyperparameters were specified a priori and were not optimized on the evaluation data, thereby reducing the risk of overfitting and enabling fair comparison across models and feature representations.

\subsection{Evaluation}
Model performance was evaluated using patient-level validation strategies designed to prevent data leakage across recordings from the same individual. Two evaluation protocols were employed: stratified group 5-fold cross-validation and leave-one-patient-out (LOPO) validation. In both approaches, all recordings from a given patient were assigned exclusively to either the training or testing set in each split. For each evaluation split, classifiers generated probabilistic predictions for individual recordings. Patient-level probabilities were computed by averaging the predicted probabilities across all recordings associated with each patient as in Eq. \ref{eq:patient_prob}, where $p_{i,j}$ denotes the predicted probability for the $j$-th recording of patient $i$, and $N_i$ denotes the total number of recordings for patient $i$. The aggregated patient-level probability $\hat{p}_i$ was calculated as:

\begin{equation}
\label{eq:patient_prob}
\hat{p}_i = \frac{1}{N_i}\sum_{j=1}^{N_i} p_{i,j}.
\end{equation}

A binary patient-level prediction was then obtained by thresholding $\hat{p}_i$ at 0.5.

Performance was assessed at the patient level using Accuracy, Sensitivity, Specificity, and F1 score as defined in Eqs.~\ref{eq:accuracy}, \ref{eq:sensitivity}, \ref{eq:specificity}, and \ref{eq:f1}, respectively. Let $TP$, $TN$, $FP$, and $FN$ denote the numbers of true positives, true negatives, false positives, and false negatives, respectively. This patient-centric evaluation framework reflects the intended clinical use case, in which diagnostic decisions are made at the patient level rather than on individual auscultation recordings.

\begin{equation}
\label{eq:accuracy}
\text{Accuracy} = {\frac{TP + TN}{TP + TN + FP + FN}}.
\end{equation}

\begin{equation}
\label{eq:sensitivity}
\text{Sensitivity} = {\frac{TP}{TP + FN}}.
\end{equation}

\begin{equation}
\label{eq:specificity}
\text{Specificity} = {\frac{TN}{TN + FP}}.
\end{equation}

\begin{equation}
\label{eq:f1}
\text{F1 Score} = {\frac{2 \cdot TP}{2 \cdot TP + FP + FN}}.
\end{equation}

 % \vspace{0.5cm }

\section{Results}
Table \ref{tab:patient_results} summarizes patient-level pediatric asthma detection performance for the top two classifiers using features extracted from pretrained self-supervised speech models under stratified group 5-fold cross-validation and LOPO validation, while incorporating sex and age. Across all feature extractors, the Wav2Vec 2.0 – HistGB model achieved the best overall performance, yielding an accuracy of 0.84, sensitivity of 0.80, specificity of 0.86, and F1-score of 0.76 under both evaluation protocols. Performance consistency between 5-fold grouped cross-validation and LOPO validation suggests promising generalization at the patient level.
Models based on WavLM features showed moderate performance, with HistGB outperforming LogReg across both validation schemes. HuBERT-based models demonstrated comparatively lower sensitivity and F1-scores, particularly under 5-fold grouped cross-validation, and consistently underperformed relative to Wav2Vec 2.0 and WavLM.

\begin{table*}[htbp]
\caption{Patient-level classification performance for pediatric asthma detection using 5-fold grouped cross-validation and LOPO validation while including sex and age.}
\begin{center}
\begin{tabular}{|l|c|c|c|c|c|c|c|c|c|}
\hline
\textbf{Feature Extractor} & \textbf{Classifier} &
\multicolumn{4}{c|}{\textbf{5-Fold grouped cross-validation}} &
\multicolumn{4}{c|}{\textbf{LOPO}} \\
\cline{3-10}
 &  &
\textbf{Accuracy} & \textbf{Sensitivity} & \textbf{Specificity} & \textbf{F1} &
\textbf{Accuracy} & \textbf{Sensitivity} & \textbf{Specificity} & \textbf{F1} \\
\hline
\multirow{2}{*}{WavLM}
& LogReg & 0.67  & 0.80 & 0.62 & 0.61 & 0.61 & 0.60 & 0.62 & 0.50 \\
& HistGB & 0.77  & 0.70 & 0.81 & 0.67 & 0.81 & 0.70 & 0.86 & 0.70 \\

\hline
\multirow{2}{*}{HuBERT}
& LogReg & 0.64 & 0.60 & 0.67 & 0.52 & 0.74 & 0.60 & 0.81 & 0.60 \\
& HistGB & 0.71 & 0.40 & 0.85 & 0.47 & 0.74 & 0.60 & 0.81 & 0.60 \\

\hline
\multirow{2}{*}{Wav2Vec 2.0}
& LogReg & 0.68 & 0.50 & 0.76 & 0.50 & 0.61 & 0.60 & 0.62 & 0.50 \\
& HistGB & \textbf{0.84} & \textbf{0.80} & \textbf{0.86} & \textbf{0.76} 
& \textbf{0.84} & \textbf{0.80} & \textbf{0.86} & \textbf{0.76} \\

\hline
\end{tabular}
\label{tab:patient_results}
\end{center}
\end{table*}

 % \vspace{0.5cm }
\section{Discussion}
To the best of our knowledge, this study is the first to evaluate the effectiveness of pretrained self-supervised speech representation models for patient-level pediatric asthma detection from breath sound recordings under clinically realistic evaluation protocols. Among the evaluated approaches, Wav2Vec 2.0 combined with HistGB consistently achieved the strongest performance, outperforming both HuBERT- and WavLM-based models across all metrics.

Importantly, this work addresses several limitations of prior asthma detection studies. Previous approaches have often relied on structured clinical measurements, such as pulmonary function tests, which are difficult to obtain in the ED and frequently infeasible in pediatric populations \cite{12}. Audio-based methods have largely been evaluated using recordings collected in controlled or non-acute environments, limiting their applicability to emergency care \cite{m11}. To the best of our knowledge, no prior study has examined pediatric asthma detection using ML and SSL-based audio representations derived from breath sounds recorded in the noisy and time-constrained ED setting. By focusing on pediatric patients, this study provides evidence that objective, audio-based asthma assessment is feasible even in challenging acute care environments.

The strong performance of Wav2Vec 2.0 is notable given its comparatively simpler pretraining objective relative to more recent models such as HuBERT and WavLM. While HuBERT and WavLM incorporate increasingly sophisticated mechanisms to model linguistic structure and environmental variability, Wav2Vec 2.0 may preserve lower-level acoustic cues that are particularly relevant for respiratory sound analysis, such as airflow turbulence, wheeze characteristics, and broadband spectral patterns. These features may be more directly informative for asthma detection than higher-level speech-oriented representations.

The consistency of Wav2Vec 2.0 – HistGB performance across both stratified group 5-fold cross-validation and the more stringent LOPO validation is clinically significant. LOPO evaluation ensures that all recordings from a given patient are excluded from training during testing, thereby eliminating patient-level data leakage and providing a realistic estimate of generalization to unseen individuals. The comparable results observed under both protocols suggest that the model is not overly reliant on patient-specific acoustic signatures and may generalize effectively in real-world pediatric settings.
In contrast, the reduced sensitivity observed for HuBERT-based models, particularly under grouped cross-validation, suggests that these representations may be less effective at capturing subtle respiratory abnormalities in noisy or heterogeneous clinical recordings. Similarly, logistic regression consistently underperformed relative to histogram-based gradient boosting, highlighting the importance of nonlinear decision boundaries when modeling complex acoustic patterns in pediatric breath sounds.

Despite these promising results, several limitations should be acknowledged. The cohort size was limited to 31 patients, which constrains statistical power and may limit the generalizability of the findings. Additionally, breath sound recordings were analyzed in isolation and did not incorporate complementary clinical information such as symptom history, which could further enhance diagnostic performance. Future work will focus on expanding the dataset, exploring multimodal integration, and assessing model robustness across different clinical environments and recording conditions.
Overall, the findings suggest that audio representations extracted from pretrained self-supervised speech models, particularly Wav2Vec 2.0, combined with nonlinear machine learning classifiers offer a viable and scalable approach for pediatric asthma detection, with potential applications in emergency and outpatient clinical settings where objective respiratory assessment remains challenging.

\section{Conclusion}
This study investigated the use of pretrained self-supervised audio representations for patient-level pediatric asthma detection from breath sound recordings collected in real-world emergency department settings. Among the evaluated approaches, Wav2Vec 2.0 combined with HistGB consistently achieved the strongest and most stable performance, providing preliminary evidence of generalization under both grouped cross-validation and LOPO evaluation. These findings highlight the potential of audio representations extracted from self-supervised speech models for pediatric asthma assessment. Future work will focus on expanding the cohort, integrating additional clinical modalities, and validating the approach across diverse clinical settings.

 % \vspace{0.5cm }
\section*{Acknowledgment}

The authors gratefully acknowledge funding support from the North York General Foundation Innovation Fund. The authors also thank Dr. Adam Hutchinson-Jaffe and Dr. Erica Merman.

 % \vspace{0.5cm }

% \balance
\bibliographystyle{ieeetr}
\bibliography{ref}
\vspace{12pt}

\end{document}